\documentclass{aa}
\usepackage[utf8]{inputenc}
\usepackage{amsmath}
\usepackage{txfonts}
\usepackage{graphics}
\usepackage{natbib}
\usepackage{float}
\usepackage{bm}
\usepackage{amssymb}
\usepackage{color}
\usepackage{epstopdf}

\renewcommand{\rm}{\textrm}
\begin{document}
\title{Onset of the Kelvin-Helmholtz instability in partially ionized magnetic flux tubes}

\author{D. Martínez-Gómez
	\and R. Soler
	\and J. Terradas}

\institute{Departament de Física, Universitat de les Illes Balears, E-07122, Palma de Mallorca, Spain \\
\email{david.martinez@uib.es}}

\abstract{Recent observations of solar prominences show the presence of turbulent flows that may be caused by Kelvin-Helmholtz instabilites (KHI). However, the observed flow velocities are below the classical threshold for the onset of KHI in fully ionized plasmas.}{We investigate the effect of partial ionization on the onset of KHI in dense and cool cylindrical magnetic flux tubes surrounded by a hotter and lighter environment.}{The linearized governing equations of a partially ionized two-fluid plasma are used to describe the behavior of small-amplitude perturbations superimposed on a magnetic tube with longitudinal mass flow. A normal mode analysis is performed to obtain the dispersion relation for linear incompressible waves. We focus on the appearance of unstable solutions and study the dependence of their growth rates on various physical parameters. An analytical approximation of the KHI linear growth rate for slow flows and strong ion-neutral coupling is obtained. An application to solar prominence threads is given.}{The presence of a neutral component in a plasma may contribute to the onset of the KHI even for sub-Alfvénic longitudinal shear flows. Collisions between ions and neutrals reduce the growth rates of the unstable perturbations but cannot completely suppress the instability.}{Turbulent flows in solar prominences with sub-Alfvénic flow velocities may be interpreted as consequences of KHI in partially ionized plasmas.}


\maketitle

\section{Introduction}\label{sec:intro}
Recent observations of the solar atmosphere have shown the presence of turbulent flows in quiescent prominences \citep[see][]{2010ApJ...716.1288B,2010SoPh..267...75R}. These phenomena have been interpreted in terms of the Rayleigh-Taylor instability (RTI) and the Kelvin-Helmhotz instability (KHI). The latter is a well-known hydrodynamic instability caused by a shear flow velocity at the interface between two fluids \citep[see][]{1961hhs..book.....C}. It is possible to find in the literature a great number of papers devoted to the study of this instability in many astrophysical environments, such as Earth's aurora \citep{1970P&SS...18.1735H}, protoplanetary disks \citep{2006ApJ...641.1131M}, the magnetopause \citep{2010JGRA..11510218G} or planetary magnetospheres \citep{1989JGR....9415113O}. In solar coronal plasmas this instability has been observed, e.g., in coronal mass ejections \citep{2011ApJ...729L...8F}.
  
Classical magnetohydrodynamic studies \citep[see, e.g.,][]{1961hhs..book.....C} have shown that, due to the effect of magnetic field, fully ionized incompressible plasmas are stable to small amplitude perturbations if the velocity of the shear flow is sub-Alfvénic. Accordingly, the magnetohydrodynamic KHI can only be triggered by super-Alfvénic shear flows. Some turbulent flows detected in quiescent prominences exhibit a behavior that resembles the non-linear stage of the KHI, but the measured velocities, lower than $30 \ \rm{km} \ \rm{s}^{-1}$ \citep{1998Natur.396..440Z,2010ApJ...716.1288B}, are below the threshold, $\gtrsim 100 \ \rm{km} \ \rm{s}^{-1}$ \citep[see][]{2008ApJ...678L.153T}, to trigger this instability. So it would seem that those turbulences cannot be interpreted as consequences of KH instabilities. However, the condition mentioned in the previous lines only applies to fully ionized plasmas, and quiescent prominences are not fully ionized but partially ionized, i.e., they are also composed of neutral particles that do not feel the magnetic field and therefore ignore its stabilizing effect. The existence of this neutral component may modify the criterion for the appearance of the KHI, allowing the onset of the instability even for sub-Alfvénic velocities.
  
The KHI in partially ionized incompressible plasmas has been studied by, e.g., \citet{2004ApJ...608..274W} and \citet{2012ApJ...749..163S} and it has been found that neutrals are unstable even for sub-Alfvénic flows. Therefore, in the absence of certain stabilizing factors, like, e.g., surface tension, partially ionized incompressible plasmas are always unstable under the presence of a velocity shear. But these results have been obtained for the case of a Cartesian interface. However, the magnetic field in the solar atmosphere is better represented by means of flux tubes. Properties of waves in a fully ionized magnetic flux tube have been investigated by, e.g., \citet{1983SoPh...88..179E} and \citet{2009A&A...503..213G}. The effect of partial ionization in a cylindrical geometry has been studied by \citet{2009ApJ...699.1553S,2013A&A...551A..86S}. However, mass flow was not included in the works by \citet{2009ApJ...699.1553S,2013A&A...551A..86S}, a necessary feature for the possible appearance of a KHI. Hence, the logical step forward is to include that previously ignored effect.
  
  In the present work we use a multi-fluid theory to examine the influence of partial ionization on the onset of KHI in magnetic flux tubes cooler and denser than their environment. More precisely, we use a two-fluid approximation \citep[see][]{1982soma.book.....P,2011A&A...529A..82Z} that treats electrons and ions as a single fluid that interacts with the other component of the plasma, the neutral fluid, by means of collisions. We include a mass flow in the longitudinal direction of the flux tube that has a discontinuity at the interface that separates the two media of different densities. In order to avoid further complexity in the model, we ignore effects like the surface tension of the fluids and compressibility. Restricting ourselves to the linear regime, we superimpose small-amplitude perturbations to the equilibrium state and derive a dispersion relation for the incompressible waves generated by those perturbations. This dispersion relation is a generalization of the formulas found in \citet{1983SoPh...88..179E} for the fully ionized case.

  This paper is organized as follows: in Section 2 we describe our model and then present the basic governing equations. In Section 3 we derive the dispersion relation for linear incompressible waves and obtain an analytical approximation to the unstable solution for slow, sub-Alfvénic flows and strong ion-neutral coupling. In Section 4 we perform a parametric study of the solutions to the dispersion relation and obtain the dependence of the KHI growth rate on the model parameters. In Section 5 we perform an application of the theory to a solar prominence thread. Finally, the conclusions of this work are given in Section 6.
  
\section{Model and equations} \label{sec:model}

  \subsection{Equilibrium state} \label{sec:equi_state}
\begin{figure} % [!ht]
  	\centering
	  	\resizebox{\linewidth}{!}{\includegraphics{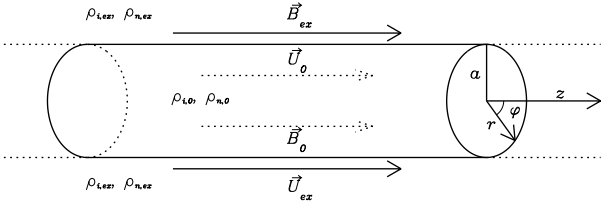}}
	  	\caption{Sketch of the model used in this work} \label{fig:model}
  \end{figure}

  The equilibrium state is a partially ionized cylindrical magnetic flux tube of radius $a$ embedded in an unbounded medium. We use cylindrical coordinates, namely $r$, $\varphi$ and $z$ for the radial, azimuthal, and longitudinal coordinates, respectively. A sketch of the model can be found in Fig. \ref{fig:model}. The subscripts $'\textrm{0}'$ and $'\textrm{ex}'$ denote quantities related to the internal and external plasmas, respectively. The densities of ions and neutrals are $\rho_{\textrm{i}}$ and $\rho_{\textrm{n}}$, respectively, and depend only on the radial direction as 
  \begin{equation} \label{eq:rho_eq1}
  	\rho_{\textrm{i}}(r)=\left\{
  	\begin{array}{l l}
  	\rho_{\textrm{i,0}} & \quad \text{if $r \leq a$}, \\
  	\rho_{\textrm{i,ex}} & \quad \text{if $r > a$},
  	\end{array} \right.
  \end{equation}
  \begin{equation} \label{eq:rho_eq2} 
  	\rho_{\textrm{n}}(r)=\left\{
  	\begin{array}{l l}
  	\rho_{\textrm{n,0}} & \quad \text{if $r \leq a$}, \\
  	\rho_{\textrm{n,ex}} & \quad \text{if $r > a$}.
  	\end{array} \right.
  \end{equation}
  \noindent Hence, there is an abrupt jump in density between the internal and external plasmas. We have chosen the particular case when the internal plasma is denser than the external one. The magnetic field, denoted by $B$, is constant and pointing along the flux tube axis, with the same value in both media, i.e., $B_{\rm{0}}=B_{\rm{ex}}$. In addition, we consider a longitudinal mass flow with constant velocity denoted by $U$. The flow velocity is discontinuous at the boundary of the flux tube.
  
  \subsection{Governing equations} \label{sec:equations}
  In this work we study  the behavior of a partially ionized plasma using a two-fluid theory. We assume the plasma is composed of an ionized fluid made of ions and electrons, and a neutral fluid made of neutral particles. The two fluids interact by means of ion-neutral collisions. The general two-fluid equations can be found in, e.g., \citet{1982soma.book.....P}, \citet{2011A&A...529A..82Z} and \citet{2014PhPl...21i2901K}. Here, we restrict ourselves to the linearized version of those equations, which describe the evolution of small-amplitude perturbations. Hence, the set of two-fluid equations that describe the behavior of linear incompressible perturbations superimposed on the equilibrium state are
  \begin{equation} \label{eq:mom_ion}
    \rho_{\rm{i}} \left(\frac{\partial}{\partial t}+U\frac{\partial}{\partial z}\right)\bm{v_{\rm{i}}} =
    -\nabla p_{\textrm{ie}}+\frac1{\mu}\left(\nabla \times \bm{b} \right)\times \bm{B}-\rho_{\rm{n}}\nu_{\textrm{ni}}
    \left(\bm{v_{\textrm{i}}}-\bm{v_{\textrm{n}}}\right),
  \end{equation}
  
  \begin{equation}\label{eq:mom_neu}
    \rho_{\textrm{n}} \left(\frac{\partial}{\partial t}+U\frac{\partial}{\partial z}\right)\bm{v_{\textrm{n}}} =
    -\nabla p_{\textrm{n}}-\rho_{\textrm{n}} \nu_{\textrm{ni}} \left( \bm{v_{\textrm{n}}}-\bm{v_{\textrm{i}}} \right),
  \end{equation}

  \begin{equation}\label{eq:induction}
    \left(\frac{\partial}{\partial t}+U\frac{\partial}{\partial z}\right)\bm{b}=\nabla \times 
    \left(\bm{v_{\textrm{i}}}\times \bm{B}\right),
  \end{equation}
  
  \begin{equation}\label{eq:incompressible}
  \nabla \cdot \bm{v_{\textrm{i}}}=\nabla \cdot \bm{v_{\textrm{n}}}=0.
  \end{equation}

  In these equations, $\bm{v_{\textrm{i}}}$ and $\bm{v_{\textrm{n}}}$ are the velocities of ions and neutrals, respectively, $p_{\textrm{ie}}$ and $p_{\textrm{n}}$ are the pressure perturbations of ion-electrons and neutrals, respectively, $\bm{b}$ is the magnetic field perturbation, $\mu$ is the magnetic permeability, $\gamma$ is the adiabatic index and $\nu_{\textrm{ni}}$ is the neutral-ion collision frequency. In addition, we define the ionization fraction as $\chi=\rho_{\textrm{n}}/\rho_{\textrm{i}}$. 
  
  In the following calculations, we replace the velocities of ions and neutrals by their corresponding Lagrangian displacements, $\bm{\xi_{\rm{i}}}$ and $\bm{\xi_{\rm{n}}}$, given by
 \begin{equation}\label{eq:Lagr_disp3}
 \bm{v_{\rm{i}}}=\frac{\partial \bm{\xi_{\rm{i}}}}{\partial t}+U\frac{\partial \bm{\xi_{\rm{i}}}}{\partial z},
 \end{equation}
 
 \begin{equation}\label{eq:Lagr_disp4}
 \bm{v_{\rm{n}}}=\frac{\partial \bm{\xi_{\rm{n}}}}{\partial t}+U\frac{\partial \bm{\xi_{\rm{n}}}}{\partial z}.
 \end{equation}

  \noindent In addition, we define the total (thermal + magnetic) pressure perturbation of the ionized fluid, $P'$, as
  \begin{equation}\label{eq:pres_tot}
    P'=p_{\rm{ie}}+\frac{\bm{B}\cdot \bm{b}}{\mu}=p_{\rm{ie}}+\frac{B b_{\rm{z}}}{\mu}.
  \end{equation}

\section{Normal mode analysis} \label{sec:normal_modes}
  From here on, we follow the same procedure as in \citet{2013A&A...551A..86S} and perform a normal mode analysis. Since the equilibrium is uniform in the azimuthal and longitudinal directions, we express the perturbations as proportional to $\exp(i m \varphi+ik_{z}z)$, where $m$ and $k_{z}$ are the azimuthal and longitudinal wavenumbers, respectively. We only retain the dependence of the perturbations on the radial direction. Furthermore, the temporal dependence is set as $\exp(-i\omega t)$, where $\omega$ is the angular frequency. Combining equations \eqref{eq:mom_ion}-\eqref{eq:induction} we arrive at a system of four coupled equations for the radial components of the Lagrangian displacements, $\xi_{\rm{r,i}}$ and $\xi_{\rm{r,n}}$, and the pressures, $P'$ and $p_{\rm{n}}$, namely 
\begin{equation} \label{eq:pre_tot}
  \frac{\partial P'}{\partial r}=\rho_{\rm{i}}\left( \Omega^2-\omega_{A}^2+i\chi \nu_{\rm{ni}}\Omega \right)
  \xi_{\rm{r,i}}-i\rho_{\rm{n}}\nu_{\rm{ni}}\Omega \xi_{\rm{r,n}},
\end{equation}

\begin{equation} \label{eq:pre_n}
  \frac{\partial p_{\rm{n}}}{\partial r}=-i\rho_{\rm{n}}\nu_{\rm{ni}}\Omega \xi_{\rm{r,i}}+\rho_{\rm{n}} \Omega(\Omega+i
  \nu_{\rm{ni}})\xi_{\rm{r,n}},
\end{equation}

\begin{equation} \label{eq:xi_ri}
  \rho_{\rm{i}}\left(\widetilde{\Omega}^2-\omega_{\rm{A}}^2\right)\frac1{r}\frac{\partial(r\xi_{\rm{r,i}})}{\partial r}=\left(\frac{m^2}{r^2}+k_{z}^2\right)\left(P'+i\frac{\nu_{\rm{ni}}}{\Omega+i
  \nu_{\rm{ni}}}p_{\rm{n}}\right),
\end{equation}

\begin{align} \label{eq:xi_rn}
  \rho_{\rm{n}}\rho_{\rm{i}}\left(\widetilde{\Omega}^2-\omega_{\rm{A}}^2\right)\frac1{r}\frac{\partial(r\xi_{\rm{r,n}})}{\partial r}=i\frac{\nu_{\rm{ni}}}
  {\Omega+i\nu_{\rm{ni}}}\rho_{\rm{n}}\left(\frac{m^2}{r^2}+k_{z}^2\right) P' \nonumber \\ -\left(\frac{m^2}{r^2}+k_{z}^2\right)\left[\frac{\rho_{\rm{n}}\nu_{\rm{ni}}^2}{(\Omega+i\nu_{\rm{ni}})^2}-\frac{\rho_{\rm{i}}\left(\widetilde{\Omega}^2-\omega_{\rm{A}}^2\right)}{\Omega(\Omega+i\nu_{\rm{ni}})}\right]p_{\rm{n}},
\end{align}

  \noindent where $\Omega=\omega - Uk_{z}$ is the Doppler-shifted frequency. Other parameters that appear in the equations are the square of the modified frequency, 
  $\widetilde{\Omega}^2$, and the square of the Alfvén frequency, $\omega_{\rm{A}}^2$, defined as
\begin{equation} \label{eq:Omega_tilde}
  \widetilde{\Omega}^2=\Omega^2\left(1+\frac{i\chi\nu_{\rm{ni}}}{\Omega+i\nu_{\rm{ni}}}\right)
\end{equation}
and
\begin{equation} \label{eq:Omega_A}
  \omega_{\rm{A}}^2=k_{z}^2c_{\rm{A}}^2,
\end{equation}
\noindent where $c_{\rm{A}}^2$ is the square of the Alfvén speed, computed as
\begin{equation} \label{eq:alfvén_speed}
c_{\rm{A}}^2=\frac{B^2}{\mu \rho_{\rm{i}}}.
\end{equation}
  Now we combine the equations \eqref{eq:pre_tot}-\eqref{eq:xi_rn} and arrive at two uncoupled equations for the pressures, namely
\begin{equation} \label{eq:pre_tot2}
  \frac{\partial^2 P'}{\partial r^2}+\frac1{r}\frac{\partial P'}{\partial r}-\left(k_{z}^2+\frac{
m^2}{r^2}\right)P'=0,
\end{equation}

\begin{equation} \label{eq:pre_n2}
  \frac{\partial^2 p_{n}}{\partial r^2}+\frac1{r}\frac{\partial p_{\rm{n}}}{\partial r}-\left(k_{z}^2+
  \frac{m^2}{r^2}\right)p_{\rm{n}}=0,
\end{equation}

  \noindent whose solutions are combinations of modified Bessel functions of the first and second kind, $I_{\rm{m}}(k_{z}r)$ and $K_{\rm{m}}(k_{z}r)$, respectively. We require the solutions to be regular at $r=0$ and vanishing at $r\to \infty$. Hence,
%\begin{multicols}{2}
    \begin{equation}\label{eq:pre_Bessels_t}
    P'(r)=\left\{
    \begin{array}{l l}
      A_{\rm{1}}I_{\rm{m}}(k_{z}r) & \quad \text{if $r \leq a$}, \\
      A_{\rm{2}}K_{\rm{m}}(k_{z}r) & \quad \text{if $r > a$},
    \end{array} \right.
  \end{equation}

  \begin{equation} \label{eq:pre_Bessels_n} 
    p_{\rm{n}}(r)=\left\{
    \begin{array}{l l}
      A_{\rm{3}}I_{\rm{m}}(k_{z}r) & \quad \text{if $r \leq a$}, \\
      A_{\rm{4}}K_{\rm{m}}(k_{z}r) & \quad \text{if $r > a$},
    \end{array} \right.
  \end{equation}
%\end{multicols}

  \noindent where $A_{\rm{1}}-A_{\rm{4}}$ are arbitrary constants. In turn, the radial components of the Lagrangian displacements of the two fluids are related to $P'$ and $p_{\rm{n}}$ as
  \begin{equation} \label{eq:xi_ri2}
  \xi_{\rm{r,i}}=\frac1{\rho_{\rm{i}}\left(\widetilde{\Omega}^2-\omega_{\rm{A}}^2\right)}\left(\frac{\partial P'}{
  	\partial r}+i\frac{\nu_{\rm{ni}}}{\Omega+i\nu_{\rm{ni}}}\frac{\partial p_{\rm{n}}}{\partial r}\right),
  \end{equation}

  \begin{align} \label{eq:xi_rn2}
    \xi_{\rm{r,n}}&=\left(\frac1{\rho_{\rm{n}}\Omega\left(\Omega+i\nu_{\rm{ni}}\right)}-\frac{\nu_{\rm{ni}}^2}{\left(\Omega+i\nu_{\rm{ni}}\right)^2}\frac1{\rho_{\rm{i}}\left(\widetilde{\Omega}^2-\omega_{\rm{A}}^2\right)}\right)\frac{\partial p_{\rm{n}}}{\partial r} \nonumber \\ &+i\frac{\nu_{\rm{ni}}}{\Omega+i\nu_{\rm{ni}}}\frac1{\rho_{\rm{i}}\left(\widetilde{\Omega}^2-\omega_{\rm{A}}^2\right)}\frac{\partial P'}{\partial r}.
  \end{align}	
  
  \subsection{Dispersion relation and approximate KHI growth rate} \label{sec:dispersion}
  To find the dispersion relation that describes the behavior of the waves in this system we need to impose that $P'$, $p_{\rm{n}}$, $\xi_{\rm{r,i}}$ and $\xi_{\rm{r,n}}$ are continuous at $r=a$, i.e., at the boundary of the tube. After applying the boundary conditions we get a system of algebraic equations for the constants $A_{\rm{1}}-A_{\rm{4}}$. The non-trivial solution to the system provides us with the dispersion relation, namely
\begin{gather}
  \left[ \frac{I_{\rm{m}}'(k_{z}a)}{I_{\rm{m}}(k_{z}a)}\rho_{\rm{n,ex}}\Omega_{\rm{ex}}\left(\Omega_{\rm{ex}}+i\nu_{\rm{ni,ex}}
  \right)-\frac{K_{\rm{m}}'(k_{z}a)}{K_{\rm{m}}(k_{z}a)}\rho_{\rm{n,0}}\Omega_{\rm{0}}\left(\Omega_{\rm{0}}+i\nu_{\rm{ni,0}}\right)
  \right] \nonumber \\
  \times \left[ \frac{I_{\rm{m}}'(k_{z}a)}{I_{\rm{m}}(k_{z}a)}\rho_{\rm{i,ex}}\left(\widetilde{\Omega}_{\rm{ex}}^2-
  \omega_{\rm{A,ex}}^2\right)-\frac{K_{\rm{m}}'(k_{z}a)}{K_{\rm{m}}(k_{z}a)}\rho_{\rm{i,0}}\left(\widetilde{\Omega}_{\rm{0}}^2-
  \omega_{\rm{A,0}}^2\right)\right] \nonumber \\
  +\frac{I_{\rm{m}}'(k_{z}a)}{I_{\rm{m}}(k_{z}a)}\frac{K_{\rm{m}}'(k_{z}a)}{K_{\rm{m}}(k_{z}a)}\frac{\rho_{\rm{n,0}}\rho_{\rm{n,ex}}\Omega_{\rm{0}}\Omega{\rm{ex}}}{\left(\Omega_{\rm{0}}+i\nu_{\rm{ni,0}}\right)\left(\Omega_{\rm{ex}}+i\nu_{\rm{ni,ex}}\right)} 
  \nonumber \\
  \times \left[\nu_{\rm{ni,0}}\left(\Omega_{\rm{ex}}+i\nu_{\rm{ni,ex}}\right)-\nu_{\rm{ni,ex}}\left(\Omega_{\rm{0}}+i\nu_{\rm{ni,0}}\right)\right]^2=0 \label{eq:dis_rel}
\end{gather}

  \noindent where the prime denotes the derivative of the modified Bessel function with respect to its argument. \par
  
  In order to simplify equation \eqref{eq:dis_rel}, we use the so-called thin tube (TT) approximation, i.e., we assume $k_{z}a \ll 1$. We perform an asymptotic expansion of the modified Bessel functions for small arguments and $m\neq 0$ and only keep the first term in the expansion. The resulting TT dispersion relation is
\begin{gather} 
  \resizebox{0.95\hsize}{!}{$\left[\rho_{\rm{i,0}}\left(\Omega_{\rm{0}}\left(\Omega_{\rm{0}}+i\chi_{\rm{0}}\nu_{\rm{ni,0}}\right)-\omega_{\rm{A,0}}^2\right)
  +\rho_{\rm{i,ex}}\left(\Omega_{\rm{ex}}\left(\Omega_{\rm{ex}}+i\chi_{\rm{ex}}\nu_{\rm{ni,ex}}\right)-\omega_{\rm{A,ex}}^2\right)
  \right]$} \nonumber \\
  \times \left[\rho_{\rm{n,0}}\Omega_{\rm{0}}\left(\Omega_{\rm{0}}+i\nu_{\rm{ni,0}}\right)+\rho_{\rm{n,ex}}\Omega_{\rm{ex}}\left(
  \Omega_{\rm{ex}}+i\nu_{\rm{ni,ex}}\right)\right] \nonumber \\
  +\left[\rho_{\rm{n,0}}\Omega_{\rm{0}}\nu_{\rm{ni,0}}+\rho_{\rm{n,ex}}\Omega_{\rm{ex}}\nu_{\rm{ni,ex}}\right]^2=0, \label{eq:TT_rel}
\end{gather}

  \noindent which is the same expression, with a slightly different notation, as Eq. (37) of \citet{2012ApJ...749..163S},
  obtained for the case of a Cartesian interface. Hence, the geometrical effect associated to the cylindrical magnetic tube disappears in the TT limit.
  
  Let us consider the limit when the collision frequencies go to zero, i.e., when the two fluids are uncoupled and show a completely independent behavior. If we neglect the terms associated with the ion-neutral collisions, we can recover from equation \eqref{eq:TT_rel} the dispersion relations for the classical  
  hydrodynamic and magnetohydrodynamic KHI \citep{1961hhs..book.....C}. In that case the dispersion relation for our model is given by
  \begin{equation}\label{eq:TT_rel2}
  \resizebox{0.9\hsize}{!}{$\left[\rho_{\rm{i,0}}(\Omega_{\rm{0}}^2-\omega_{\rm{A,0}}^2)+\rho_{\rm{i,ex}}(\Omega_{\rm{ex}}^2-\omega_{\rm{A,ex}}^2)\right][\rho_{\rm{n,0}}\Omega_{\rm{0}}^2+\rho_{\rm{n,ex}}\Omega_{\rm{ex}}^2]=0,$}
  \end{equation}
  
  \noindent from which we can obtain separated solutions for the ionized component of the plasma and the neutral one. On the one hand, the solutions for the neutral fluid are given by
  \begin{equation}\label{eq:sol_neu}
  \omega=k_{z}\frac{\rho_{\rm{n,0}}U_{\rm{0}}+\rho_{\rm{n,ex}}U_{\rm{ex}}}{\rho_{\rm{n,0}}+\rho_{\rm{n,ex}}}\pm i k_{z}|U_{\rm{0}}-U_{\rm{ex}}|\frac{\sqrt{\rho_{\rm{n,0}}\rho_{\rm{n,ex}}}}{\rho_{\rm{n,0}}+\rho_{\rm{n,ex}}},
  \end{equation} 
  where the branch with a positive imaginary part implies that the amplitudes of the perturbations grow with time. This growing solution exists for any value of the shear flow velocity. Therefore, the neutral fluid is always unstable in the presence of a shear flow.
  
  On the other hand, for the ionized component we have the following solutions:
  \begin{align}\label{eq:sol_ion}
  \omega&=k_{z}\frac{\rho_{\rm{i,0}}U_{\rm{0}}+\rho_{\rm{i,ex}}U_{\rm{ex}}}{\rho_{\rm{i,0}}+\rho_{\rm{i,ex}}} \nonumber \\ &\pm k_{z}\left[\frac{B_{\rm{0}}^2+B_{\rm{ex}}^2}{\mu(\rho_{\rm{i,0}}+\rho_{\rm{i,ex}})}-(U_{\rm{0}}-U_{\rm{ex}})^2\frac{\rho_{\rm{i,0}}\rho_{\rm{i,ex}}}{(\rho_{\rm{i,0}}+\rho_{\rm{i,ex}})^2}\right]^{1/2},
  \end{align}
  
  \noindent from which it can be seen that the magnetohydrodynamic KHI only appears when
  \begin{align}\label{eq:threshold}
  |U_{\rm{0}}-U_{\rm{ex}}| &> \sqrt{\frac{B_{\rm{0}}^2+B_{\rm{ex}}^2}{\mu}\frac{\rho_{\rm{i,0}}+\rho_{\rm{i,ex}}}{\rho_{\rm{i,0}}\rho_{\rm{i,ex}}}} \nonumber \\ &=\sqrt{\frac{\left(\rho_{\rm{i,0}}c_{\rm{A,0}}^2+\rho_{\rm{i,ex}}c_{\rm{A,ex}}^2\right)\left(\rho_{\rm{i,0}}+\rho_{\rm{i,ex}}\right)}{\rho_{\rm{i,0}}\rho_{\rm{i,ex}}}}.
  \end{align}
  The presence of this velocity threshold is caused by the effect of the longitudinal magnetic field on the ionized fluid. The velocity threshold is super-Alfvénic, which indicates that the typically observed flows along, e.g., prominence threads or spicules cannot trigger a KHI if the plasma is fully ionized. This same criterion for the velocity threshold can be found in other works devoted to the study of waves in fully ionized plasmas with longitudinal mass flows. For instance, equation \eqref{eq:threshold} is equivalent (if we assume $c_{\rm{A,ex}}=0$) to equation (5) of \citet{1988JETP..67..1594}.
  
  In addition, according to \citet{1988JETP..67..1594} and \citet{2010SoPh..267...75R}, a fully ionized plasma may be subject to an explosive instability if the shear flow velocity, $\Delta U$, fulfills the following condition:
  \begin{equation}\label{eq:explosive}
  c_{\rm{A,0}}\left(\frac{\rho_{\rm{i,0}}}{\rho_{\rm{i,ex}}}\right)^{1/2}<\Delta U<c_{\rm{A,0}}\left(\frac{\rho_{\rm{i,0}}+\rho_{\rm{i,ex}}}{\rho_{\rm{i,ex}}}\right)^{1/2}.
  \end{equation}
  Therefore, taking into account the suggested existence of the explosive instability, three unstable branches would be expected to appear in the set of solutions to the dispersion relation: the first one, associated to the neutrals, would be present for any value of the shear flow velocity; the second one, the explosive instability, would appear between the shear flow velocity thresholds given by equation \eqref{eq:explosive}; and the third one, associated to the ions, would follow the criterion of equation \eqref{eq:threshold}. (Note that if $B_{\rm{ex}}\ne 0$ but $B_{\rm{ex}}=B_{\rm{0}}$, as in our model, the lower and upper limits of equation \eqref{eq:explosive} should be multiplied by the factor $\sqrt{2}$).
  
  Let us go back to the coupled case. The full dispersion relation, Eq. \eqref{eq:dis_rel}, must be solved numerically when ion-neutral collisions are at work. However, it is possible to find an approximate solution when the ion-neutral coupling is strong and sub-Alfvénic flows are considered. For simplicity, we also assume that the external flow velocity is zero, i.e., $U_{\rm{ex}}=0$. For sub-Alfvénic flows the only unstable solution we obtain from the dispersion relation in the uncoupled case is that associated with neutrals, Eq. \eqref{eq:sol_neu}, since the magnetic field is able to stabilize ions, Eq. \eqref{eq:sol_ion}. Hence, we try to find a correction to the neutrals-related solution due to ion-neutral collisions. To do so, we write $\omega=\omega_{\rm{0}}+i\gamma$, where $\omega_{\rm{0}}$ is the neutrals' unstable solution given in equation \eqref{eq:sol_neu} and $\gamma$ is a small correction. We insert this expression for $\omega$ into the TT dispersion relation, Eq. \eqref{eq:TT_rel}, and only keep up to first-order terms in $\gamma$ and second-order terms in $U_{\rm{0}}$. After some algebraic manipulations we find a solution for $\gamma$. We omit the details for the sake of simplicity. Finally, the following approximate solution for the frequency is obtained,
  \begin{equation}\label{eq:approx}
  \omega \approx \frac{k_{z}U_{\rm{0}}\rho_{\rm{n,0}}}{\rho_{\rm{n,0}}+\rho_{\rm{n,ex}}} 
  +i\frac{2k_{z}^2U_{\rm{0}}^2\rho_{\rm{n,0}}\rho_{\rm{n,ex}}}{(\rho_{\rm{n,0}}+\rho_{\rm{n,ex}})(\nu_{\rm{ni,0}}\rho_{\rm{n,0}}+\nu_{\rm{ni,ex}}\rho_{\rm{n,ex}})}.
  \end{equation}
  The approximated growth rate, i.e., the imaginary part of $\omega$, is quadratic in the flow velocity and inversely proportional to the collision frequencies. Hence, if the same parameters are considered, the value of the growth rate is smaller in the strongly coupled case than in the uncoupled case (compare with Eq. \eqref{eq:sol_neu}). Therefore, ion-neutral collisions have a stabilizing effect.
  
\section{Exploring the parameter space} \label{sec:par_space}
  Due to its complexity, equation \eqref{eq:dis_rel} must be solved numerically. In 
  this section we study the dependence of the solutions of the dispersion relation with respect to various physical parameters. Also we compare the full results with the analytical approximation shown in the previous section. 
  
\begin{figure*}
  	\centering
  	\begin{tabular}{@{}ccc@{}}
  		\includegraphics[width=0.32\linewidth]{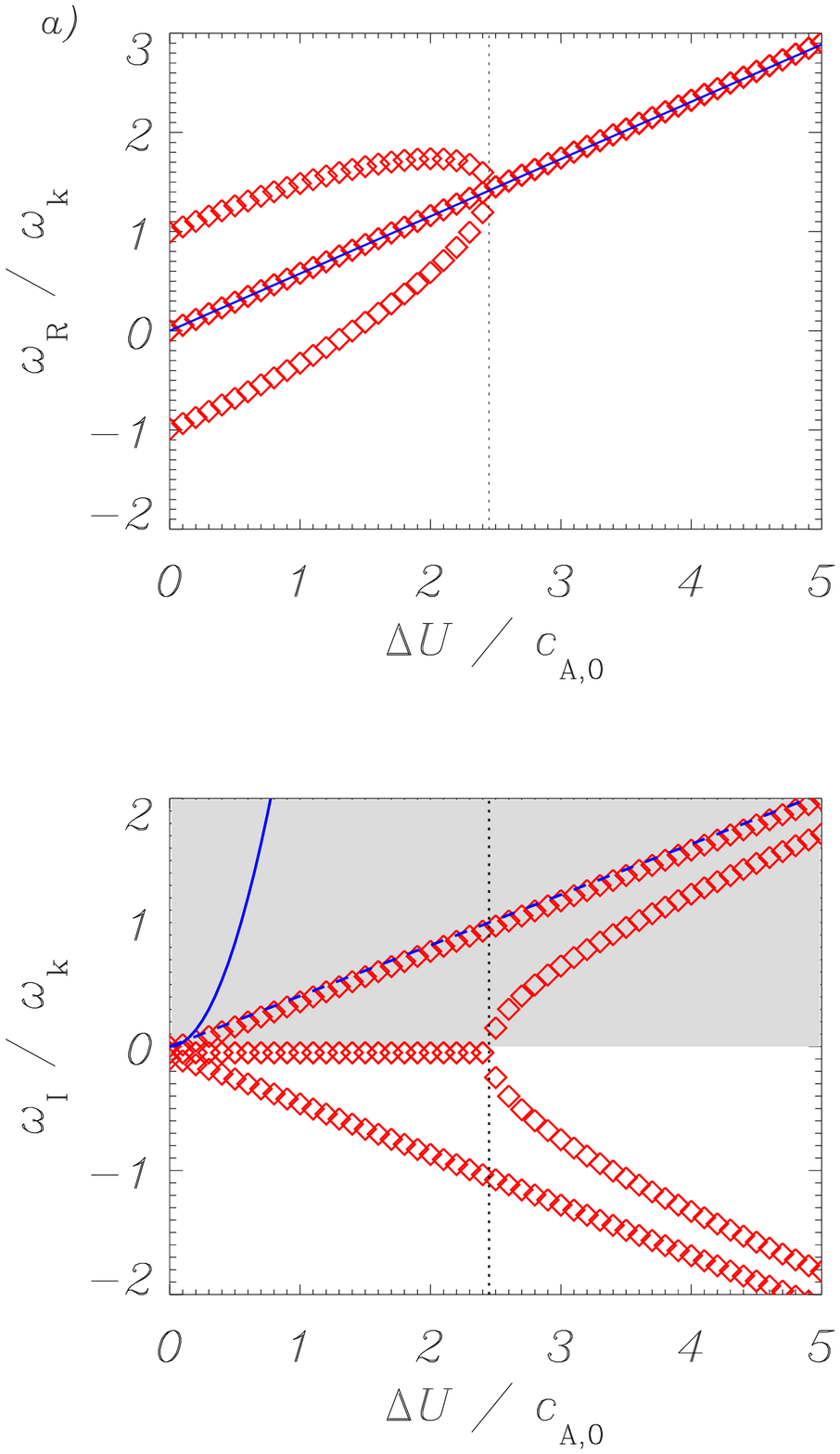} &
  		\includegraphics[width=0.32\linewidth]{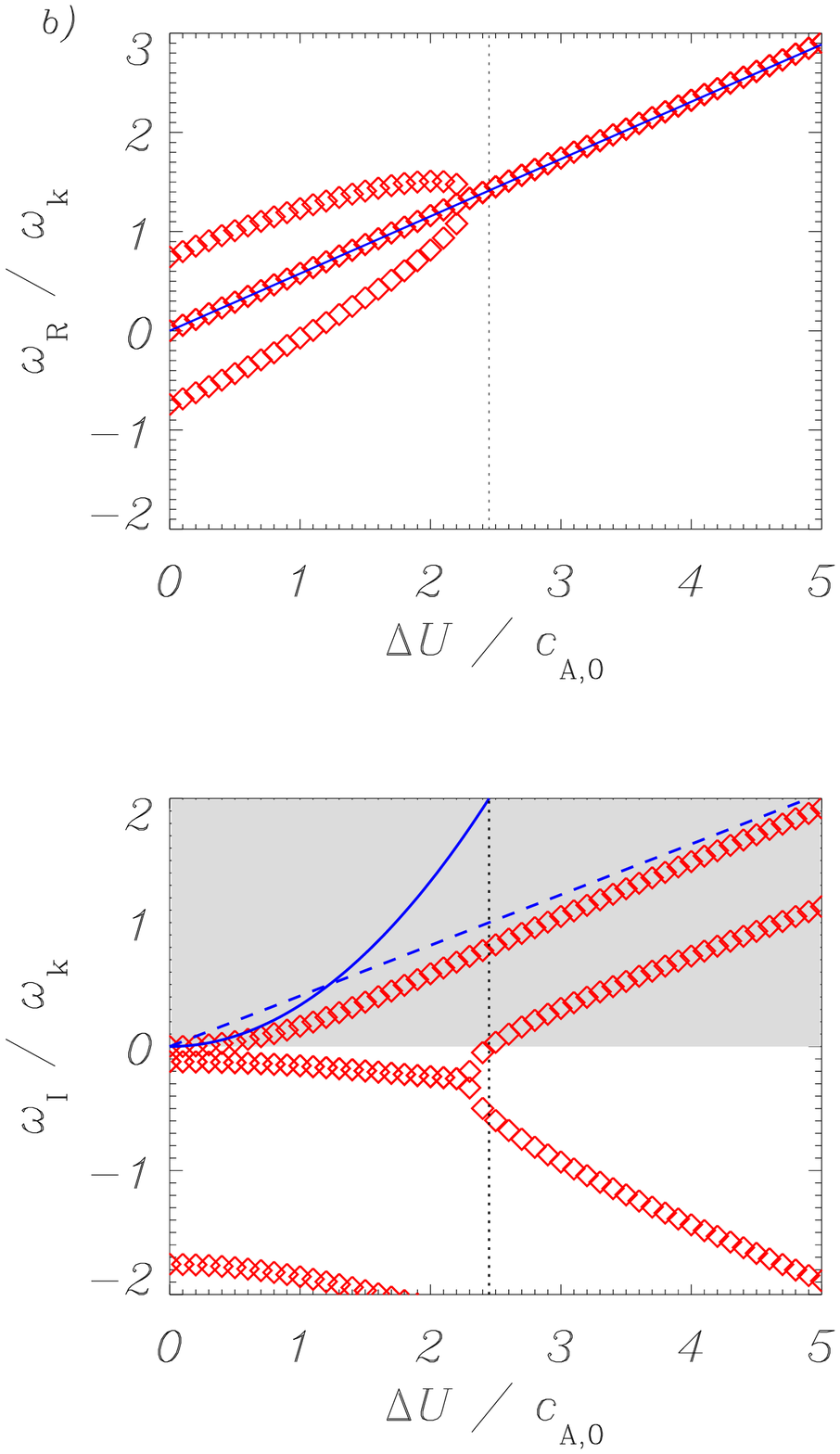} &
  		\includegraphics[width=0.32\linewidth]{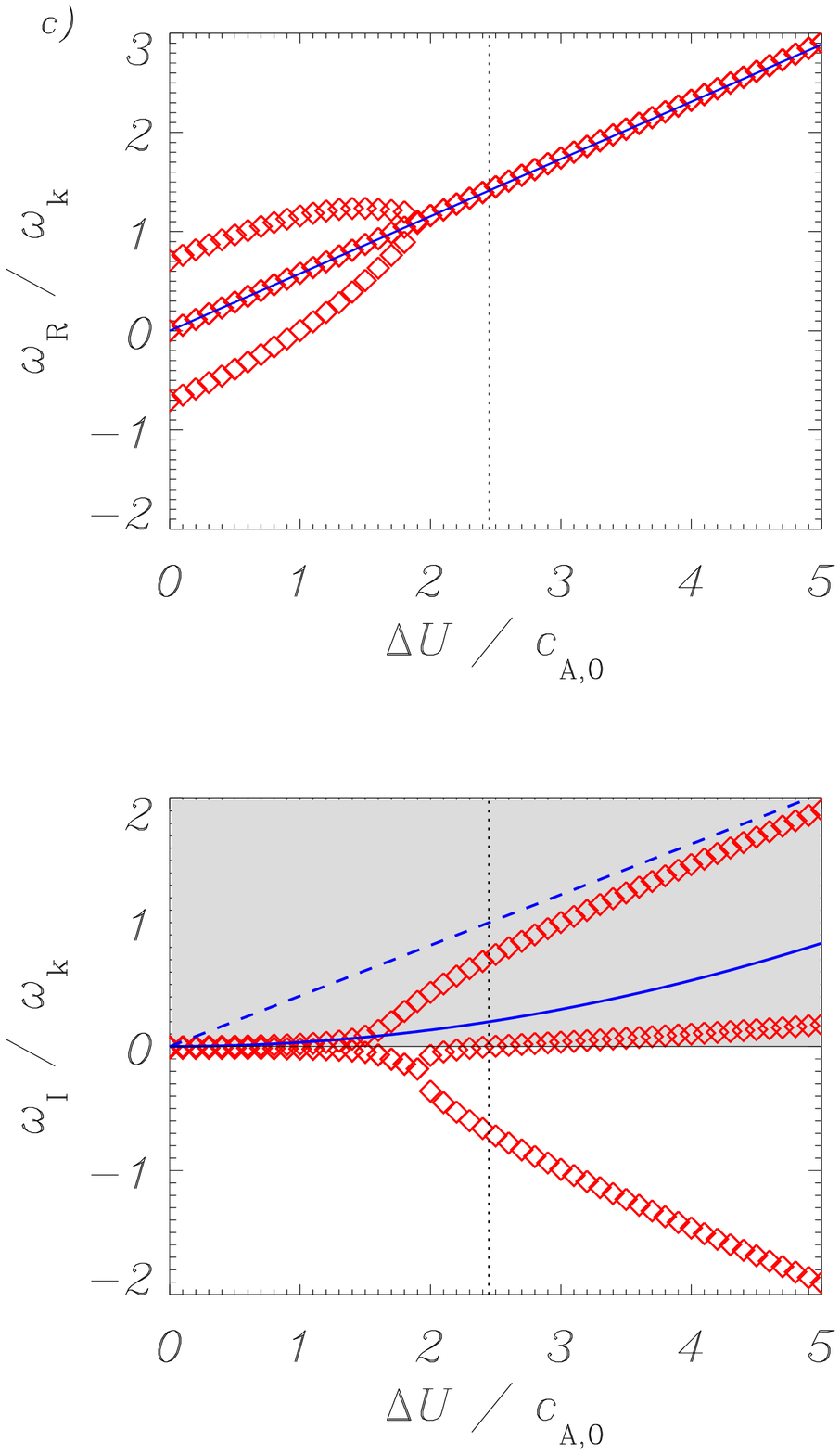} 
  	\end{tabular}
  	\caption{Upper panels: $\omega_{\rm{R}}/\omega_{\rm{k}}$ as a function of the normalized shear flow velocity, $\Delta U/c_{\rm{A,0}}$, for $k_{z}a=0.1$, $m=1$, and three different collision frequencies ($\bm{a)} \ \nu_{\rm{ni}}/\omega_{\rm{k}}=0.1$, $\bm{b)} \ \nu_{\rm{ni}}/\omega_{\rm{k}}=1$ and $\bm{c)} \ \nu_{\rm{ni}}/\omega_{\rm{k}}=10$). Lower panels: $\omega_{I}/\omega_{\rm{k}}$ as a function of $\Delta U/c_{\rm{A,0}}$ for the same set of parameters as above. The red symbols represent the solutions of the full dispersion relation, i.e., Eq. \eqref{eq:dis_rel}; the blue solid lines correspond to the analytical approximation given by equation \eqref{eq:approx} and the blue dashed lines show the unstable branch of the neutral fluid when there is no coupling (Eq. \eqref{eq:sol_neu}).}  \label{fig:deltaU}
  \end{figure*}

  Throughout this section we use dimensionless parameters. Unless otherwise stated we use $\rho_{i,0}/\rho_{i,ex}=2$, $\rho_{n,0}/\rho_{n,ex}=2$, $c_{\rm{A,0}}=1$ and $k_{z}a=0.1$. For simplicity, we assume that the collision frequency has the same value in both internal and external plasmas, so we drop the subscripts from $\nu_{\rm{ni}}$. In addition, we now focus on the kink mode, the only mode that causes displacements of the axis of the cylinder, so we use $m=1$. All frequencies have been normalized with respect to the kink frequency, $\omega_{\rm{k}}$, which is the frequency of the kink wave in the TT limit in the fully ionized case \citep[see, e.g.,][]{1976JETP...43..491R,1981A&A....98..155S}. The kink frequency is given by
  \begin{equation}\label{eq:kink}
	 \omega_{k}=k_{z}\sqrt{\frac{\rho_{i,0}c_{A,0}^2+\rho_{i,ex}c_{A,ex}^2}{\rho_{i,0}+\rho_{i,ex}}}.
  \end{equation}
  To begin with, we study the dependence of the solutions of the dispersion relation with respect to the shear flow velocity. Hence, we take $\Delta U \equiv U_{\rm{0}}-U_{\rm{ex}}$ as a free variable. We consider three different values for the collision frequency, which allow us to investigate the behavior of the solutions depending on the strength of the ion-neutral coupling. Fig. \ref{fig:deltaU} displays the real and imaginary parts of the frequency as functions of the normalized shear flow velocity for $a)$ weak coupling ($\nu_{\rm{ni}}/\omega_{\rm{k}}=0.1$), $b)$ intermediate coupling ($\nu_{\rm{ni}}/\omega_{\rm{k}}=1$) and $c)$ strong coupling ($\nu_{\rm{ni}}/\omega_{\rm{k}}=10$). The red symbols represent the solutions obtained numerically from the complete dispersion relation, Eq. \eqref{eq:dis_rel}. The full numerical results are compared with the analytical solutions in the strongly coupled limit (shown as blue solid lines) and in the uncoupled case (shown as blue dashed lines). In addition, the classical shear flow velocity threshold for the KHI in a fully ionized fluid (Eq. \eqref{eq:threshold}) is denoted by the vertical dotted lines.
  The upper panels of Fig. \ref{fig:deltaU} display the real part of the frequency. We observe a very similar behavior for the three studied cases. Initially, when the shear flow velocity is zero, we find two solutions with nonzero $\omega_{\rm{R}}$. These solutions are associated to the ionized fluid and correspond to the usual kink MHD wave found in fully ionized tubes \citep{1983SoPh...88..179E}. The solution with $\omega_{\rm{R}}>0$ is the forward propagating kink wave, while the solution with $\omega_{\rm{R}}<0$ is the backward propagating kink wave. A third solution with $\omega_{\rm{R}}>0$ emerges when the shear flow velocity increases from zero. The new solution is associated to the neutral component of the plasma, in the sense that this solution only appears in the presence of neutrals. However, note that such simple associations between solutions and fluids cannot be done when the coupling is high and ions and neutrals behave as a single fluid. As the flow velocity keeps increasing, the three solutions converge for a critical flow velocity that depends on the collision frequency. The stronger the ion-neutral coupling, the smaller the critical flow. From that point on, the real part of the frequency is proportional to $\Delta U$ and is well described by the real part of Eq. \eqref{eq:sol_neu} or, equivalently, Eq. \eqref{eq:approx}. \par
  
  The lower panels of Fig. \ref{fig:deltaU} display the imaginary part of the frequency. The differences between the panels are much more remarkable than before, meaning that the value of the collision frequency has an important effect on the imaginary part of the frequency. The shaded zone denotes the region of instability, i.e., $\omega_{\rm{I}}>0$. In such a case, perturbations exponentially grow with time (in the opposite case perturbations are damped). We observe that for small shear flow velocities there is only one unstable solution, corresponding to that originally associated to the neutral component of the plasma. A second unstable branch (originally associated to ions) appears for larger flow velocities above the classical super-Alfvénic threshold (Eq. \eqref{eq:threshold}). Importantly, the only instability present for slow, sub-Alfvénic speeds is that originally associated to the neutral component of the plasma regardless of the collisional coupling between ions and neutrals. Ion-neutral collisions reduce the growth rate of that instability to a great extent, but collisions are not able to completely suppress the instability \citep{2004ApJ...608..274W,2012ApJ...749..163S}. We have overplotted with a blue solid line the analytical approximation to the growth rate (Eq. \eqref{eq:approx}). The approximation informs us that the growth rate is directly proportional to the square of the shear flow velocity and inversely proportional to the ion-neutral collision frequency. As expected, the approximation shows a good agreement with the numerical results for small shear flow velocities. For the case of weak coupling, the approximation is reasonably good for flow velocities up to 40\% the internal Alfvén speed. When the collision frequency is increased, the range of agreement between the numerical results and the approximation is greatly extended to super-Alfvénic speeds. Finally, note that in the left panel there is a stable solution that is not present in the other two panel (see the bottom-most curve). The reason for this absence is that for high collision frequencies this solution moves out of the vertical scale used in the plots. We are not interested in this solution because it is always stable.
  
  Concerning the explosive instability hinted by \citet{1988JETP..67..1594}, for our set of parameters this phenomenon would appear in the range $2<\Delta U / c_{\rm{A,0}}<\sqrt{6}$ (when $\nu_{\rm{ni}}=0$; the existence of collisions between ions and neutrals would modify this criterion). If we pay attention to the lower panels of Fig. \ref{fig:deltaU}, we find that the only solution with positive imaginary part of the angular frequency is the one associated to the neutrals. Thus, we observe no evidence of the explosive instability. The only additional remarkable feature in that range is found in the upper panels of the same figure: the real part of the angular frequency of one of the solutions associated to ions turns from being negative to being positive, meaning a change in the direction of the propagation of the wave.
  
\begin{figure} %[!ht]
  \centering
	\resizebox{\hsize}{!}{\includegraphics[width=0.5\linewidth]{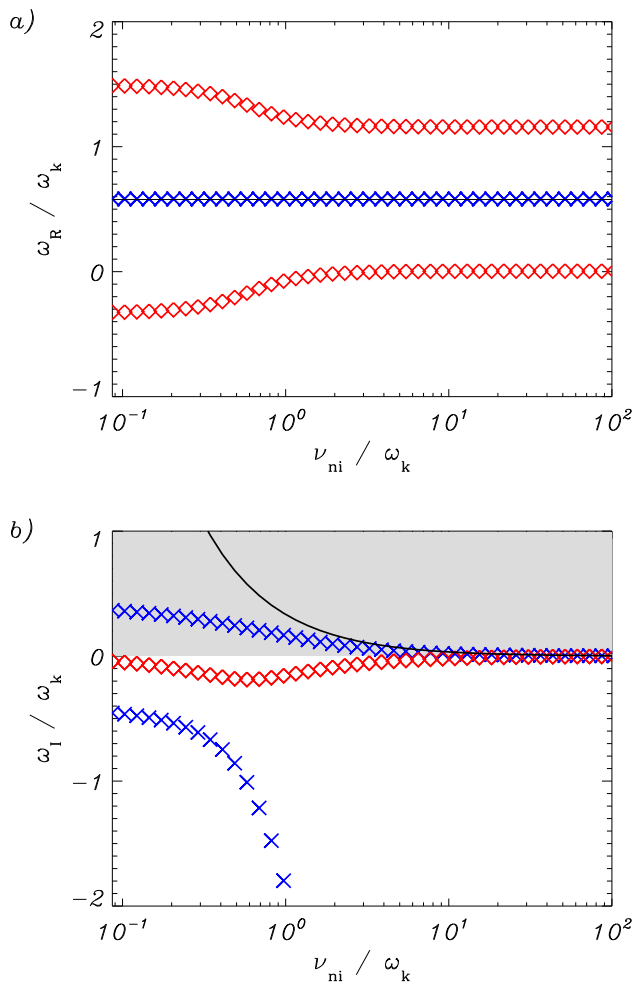}}
    \caption{ \textbf{a)} $\omega_{\rm{R}}/\omega_{\rm{k}}$ and \textbf{b)} $\omega_{\rm{I}}/\omega_{\rm{k}}$ for the kink mode (m=1) as a function of $\nu_{\rm{ni}}/\omega_{\rm{k}}$, with $\Delta U/c_{\rm{A,0}}=1$ and $k_{z}a=0.1$. The red diamonds are the solutions originally associated to the ions when there is no coupling, while the blue crosses are the solutions for neutrals. The solid line is the analytical approximation given by Eq. \eqref{eq:approx}. In \textbf{b)} the shaded area denotes the region of instability.\label{fig:col_freq}}
\end{figure}

  To investigate in more detail the effect of ion-neutral collisions on the instability for slow flows, we perform the following study. We fix the shear flow velocity to $\Delta U/c_{\rm{A,0}}=1$ and compute the frequencies as function of the ion-neutral collision frequency, $\nu_{\rm{ni}}$. The chosen flow velocity is below the classical threshold for the KHI in fully ionized plasmas \citep{1961hhs..book.....C}, so that only the neutral component is unstable in this configuration. The results are displayed in Figure \ref{fig:col_freq}, where the solutions originally associated to ions are shown as red diamonds and those originally associated to neutrals are plotted with blue crosses. There is always one unstable solution for any value of $\nu_{\rm{ni}}$, but its growth rate decreases when the collision frequency increases. The growth rate gets reduced because neutrals feel indirectly, through the collisions with ions, the stabilizing effect of the magnetic field. In addition, as discussed before, the analytical approximation for the growth rate, Eq. \eqref{eq:approx}, shows a good agreement with the numerical results for high values of the collision frequencies, as consistent with the assumptions behind the approximation. Concerning the real part of the frequency, we note that the frequency of solutions associated to ions decreases until it reaches a \textit{plateau} for $\nu_{\rm{ni}}/\omega_{\rm{k}}>1$ \citep{2013A&A...551A..86S}, while the frequency of the solution associated to the neutrals stays constant all over the range.
  
For the sake of completeness, we have also studied the behavior of the solutions depending on $m$. The mode with $m=0$ is known as the sausage mode and produces expansions and contractions of the plasma tube, without displacing its axis \citep{1983SoPh...88..179E}. Modes with $m>1$ are known as fluting modes \citep{1983SoPh...88..179E}. It must be noted that, as can be seen in Eq. \eqref{eq:TT_rel}, the TT limit is independent of the value of $m$ for $m \neq 0$; this fact implies that in the range of applicability of that approximation there will not be substantial variations in the behavior of the different modes. To observe some dissimilarities we need to choose parameters beyond the TT case. Thus, we keep the values of the parameters used in the previous section but change the dimensionless longitudinal wavenumber to a larger value, namely, $k_{z}a=2$, and take $\nu_{\rm{ni}}/\omega_{\rm{k}}=1$. The results show that there are minor differences between each solution: the solutions weakly depend on $m$ and, for large $m$, they become independent of this parameter.

We have also repeated the calculations done in this section for higher values of the longitudinal wavenumber, $k_{z}$. We have not included the results here because they are not significantly different from what we have already explained. When $k_{z}$ is increased, the system distances from the regime of the TT approximation, but this only produces a slight variation in the real part of the normalized angular frequencies, while the imaginary part, i.e., the growth rates, remain almost unaltered.

\section{Application to solar prominence threads} \label{sec:application}
In this section we perform a specific application to solar prominence plasmas. We solve the full dispersion relation, Eq. \eqref{eq:dis_rel}, with values representative of a quiescent prominence. Therefore, the internal medium represents a prominence thread with densities of ions and neutrals such that $\rho_{\rm{i,0}}+\rho_{\rm{n,0}}=10^{-9} \ \rm{kg} \ \rm{m}^{-3}$, a temperature of $T_{\rm{0}}=7000 \ \rm{K}$ and radius of $a=100 \ \rm{km}$; the external medium is composed of inter-thread plasma with $\rho_{\rm{i,ex}}+\rho_{\rm{n,ex}}=2\times 10^{-10} \ \rm{kg} \  \rm{m}^{-3}$ and $T_{\rm{ex}}=35000 \ \rm{K}$, which corresponds to the regime of prominence-corona transition region (PCTR). Densities and temperatures are chosen for the equilibrium condition of the total pressure (thermal plus magnetic) to be fulfilled, i.e., the total pressure is the same in both media. The magnetic fields are $B_{\rm{0}}=B_{\rm{ex}}=10 \ \rm{G}$. We focus on the kink mode, so we take $m=1$. The neutral-ion collision frequencies depend on the temperatures and densities and are computed using the following expression \citep[see][]{1965RvPP....1..205B}
\begin{equation}\label{eq:braginskii}
\nu_{\rm{ni}}=\frac{\rho_{\rm{i}}}{2m_{\rm{p}}}\sqrt{\frac{16k_{\rm{B}}T}{\pi m_{\rm{p}}}}\sigma_{\rm{in}},
\end{equation}
\noindent where $m_{\rm{p}}$ is the proton mass, $k_{\rm{B}}$ is the Boltzmann constant, $\sigma_{\rm{in}}\approx 5\times 10^{-19} \ \rm{m}^2$ is the collisional cross section for a hydrogen plasma.

\begin{figure*}
	\centering
	\begin{tabular}{@{}ccc@{}}
		\includegraphics[width=0.5\linewidth]{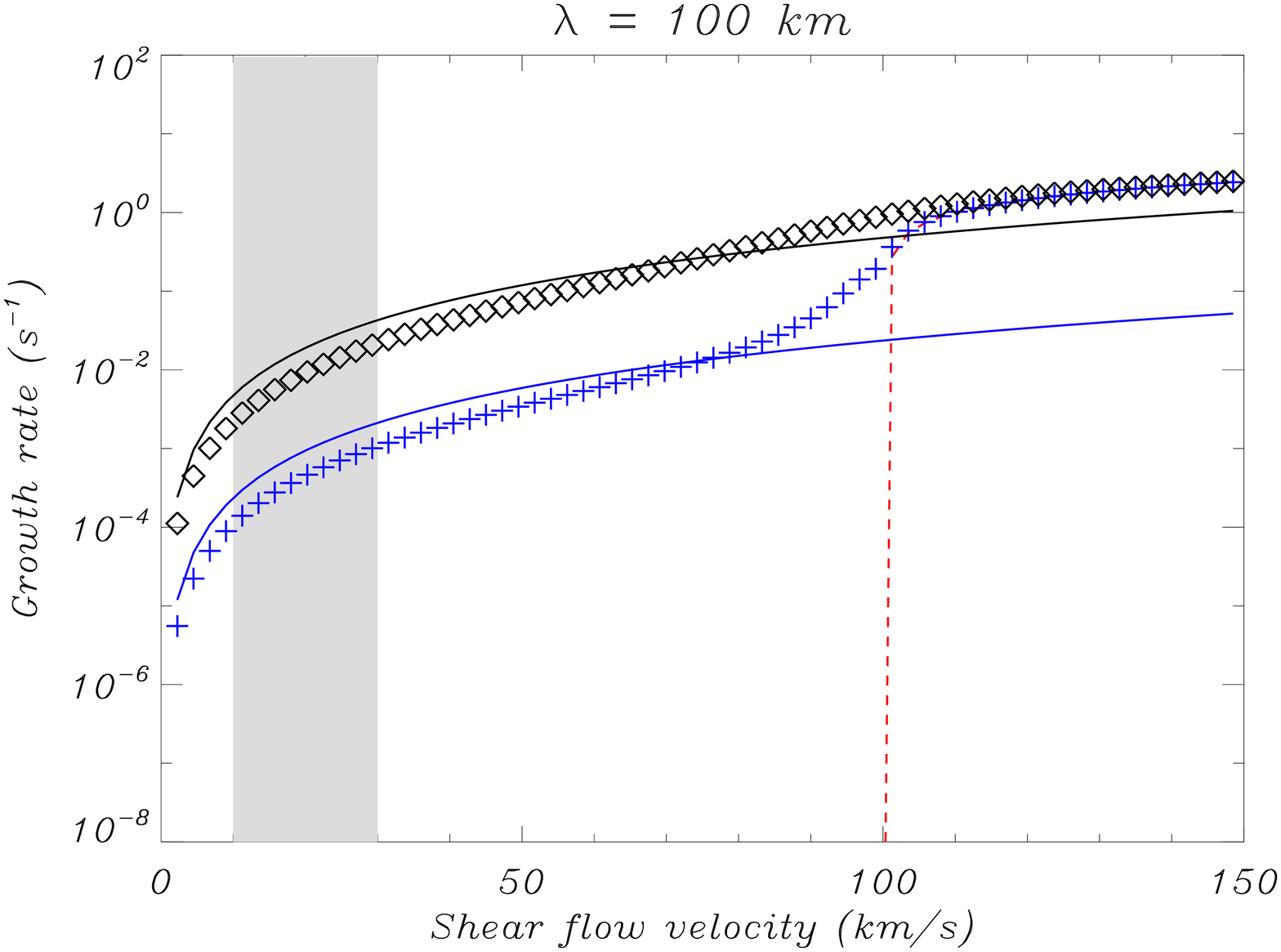} &
		\includegraphics[width=0.5\linewidth]{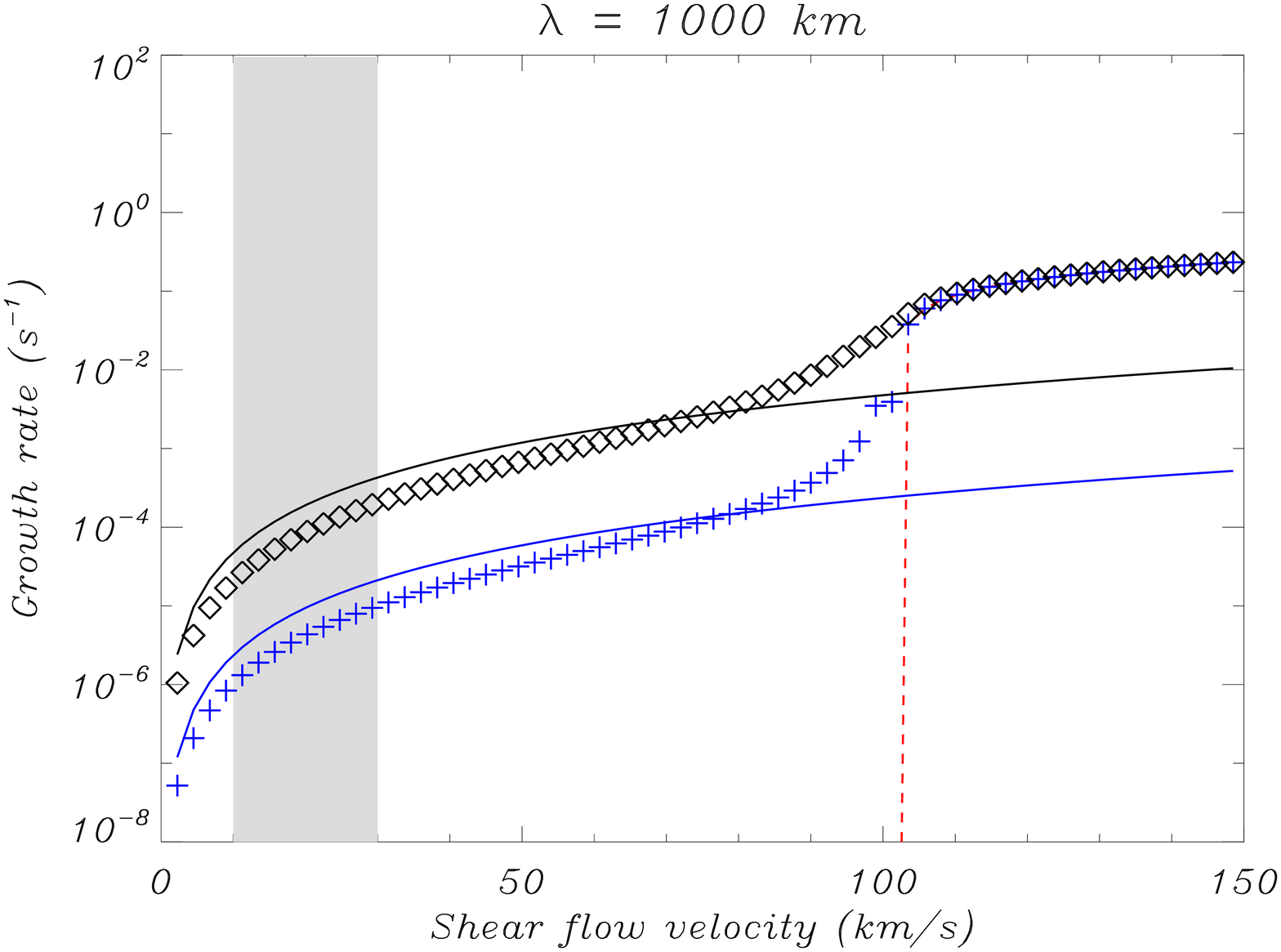} 
	\end{tabular}
	\caption{Application to solar prominence threads. Growth rates as functions of the shear flow velocity for the following set of parameters: $\rho_{\rm{i,0}}+\rho_{\rm{n,0}}=10^{-9} \ \rm{kg} \ \rm{m}^{-3}$, $\rho_{\rm{i,ex}}+\rho_{\rm{n,ex}}=2\times 10^{-10} \ \rm{kg} \ \rm{m}^{-3}$, $B_{\rm{0}}=B_{\rm{ex}}=10^{-3} \ \rm{T}$, $a=100 \ \rm{km}$, $T_{\rm{0}}=7000 \ \rm{K}$ and $T_{\rm{ex}}=35000 \ \rm{K}$; on the left panel the wavenumber is $k_{z}=2\pi/\lambda=2\pi\times 10^{-5} \ \rm{m}^{-1}$ and on the right panel $k_{z}=2\pi\times 10^{-6} \ \rm{m}^{-1}$. The red dashed lines correspond to a fully ionized plasma ($\chi=0$), the blue crosses to a partially ionized case ($\chi=4$) and the black diamonds to a weakly ionized case ($\chi=100$). The solid lines represent the solutions given by the analytical approximation in Eq. \eqref{eq:approx}. The shaded zone denotes the region of values of flow velocity that have been frequently measured in solar prominences.}  \label{fig:application}
\end{figure*}

Fig. \ref{fig:application} displays the most unstable solution of Eq. \eqref{eq:dis_rel} as a function of the shear flow velocity for three different values of the ionization fraction: the red dashed lines represent the fully ionized case ($\chi=0$), the blue crosses represent a partially ionized situation ($\chi=4$) and the black diamonds depict a weakly ionized case ($\chi=100$). The left panel shows the results for a wavelength $\lambda=100 \ \rm{km}$, that correspond to a longitudinal wavenumber $k_{z}=2\pi/\lambda=2\pi\times 10^{-5} \ \rm{m}^{-1}$. In the right panel the wavelength used is $\lambda=1000 \ \rm{km}$, so $k_{z}=2\pi\times 10^{-6} \ \rm{m}^{-1}$. The shaded zone of Fig. \ref{fig:application} denotes a range of typical velocities (from $10 \ \rm{km} \ \rm{s}^{-1}$ to $30 \ \rm{km} \ \rm{s}^{-1}$) that have been measured in quiescent prominences \citep{1998Natur.396..440Z,2010ApJ...716.1288B}. The limits of this zone could slightly vary depending on the observations that are chosen as reference, but this variation is not significant for our analysis. We see that for a fully ionized case the instability only appears for shear flow velocities far from the detected values. On the contrary, the cases with a neutral component show instabilities for all the range of velocities. Hence, partial ionization may explain the occurrence of KHI in solar prominence plasmas even when the observed flows are below the classical threshold. \par
If we compare both panels in Fig. \ref{fig:application}, we notice that on the right panel the growth rates are lower than on the left one. More precisely, they are about one order of magnitude smaller when the shear flow velocities are large. Since the wavenumber on the right panel is one order of magnitude smaller than on the left one, this behavior is in agreement with Eq. \eqref{eq:sol_neu}. On the other hand, when the velocities are low, the growth rates on the right are two orders of magnitude smaller, which is  consistent with what it is derived from Eq. \eqref{eq:approx}.

In addition, in Fig. \ref{fig:application} we have overplotted with solid lines the approximate analytical solutions given by the imaginary part of Eq. \eqref{eq:approx}. The growth rates obtained from the analytical approximation are within the same order of magnitude; thus, Eq. \eqref{eq:approx} may be used to calculate estimations of the growth rates of KHI in a real prominence thread without the need of solving the much more complex full dispersion relation. \par

Our analysis have demonstrated that KHI may be present in quiescent prominences due to the effect of partial ionization. However, we are not yet allowed to state that this magnetohydrodynamic instability may explain the observed turbulent flows. Before that it is necessary to check if the growth rates given by the theory are consistent with an instability that could be actually observed. Estimated lifetimes of prominence threads are about 20 minutes \citep{2005SolarPhysics}. For the region of highest observed velocities we compute the following growth times of the instability ($\tau\equiv 1/\omega_{\rm{I}}$): $a) \ \lambda=100 \ \rm{km} \Rightarrow \tau\sim 100 \ \rm{s}$ for the weakly ionized case ($\chi=100$) and $\tau\sim 1000 \ s$ for $\chi=4$; $b) \ \lambda=1000 \ \rm{km} \Rightarrow \tau\sim 10^{4} \ \rm{s} \ (\sim 2.8 \ \rm{hours})$ for $\chi=100$ and $\tau\sim 10^{5} \ \rm{s} \ (\sim 28 \ \rm{hours})$ for $\chi=4$. The growth times obtained for $\lambda=1000 \ \rm{km}$ are larger than the typical lifetime of a thread; therefore, an instability originated by a perturbation with this wavelength cannot be the cause of the observed turbulent flows. On the contrary, the growth times for the shortest wavelength are of the same order of magnitude or lower than the detected lifetimes. This means that during the life of a prominence thread there is enough time for the development of a KHI caused by a perturbation with that wavelength.

In addition, the analytical approximation we have derived in Section \ref{sec:dispersion}, i.e., Eq. \eqref{eq:approx}, may be a useful tool in the field of prominence seismology \citep{2014IAUS..300...30B}. We introduce a new parameter $\bar{\nu}$, called mean collision frequency of the plasma, and defined through the following relation
\begin{equation} \label{eq:mean_col}
	\frac1{\bar{\nu}}=\frac{2\rho_{\rm{n,0}}\rho_{\rm{n,ex}}}{\left(\rho_{\rm{n,0}}+\rho_{\rm{n,ex}}\right)\left(\rho_{\rm{n,0}}\nu_{\rm{ni,0}}+\rho_{\rm{n,ex}}\nu_{\rm{ni,ex}}\right)}.
\end{equation}
Thus, the imaginary part of Eq. \eqref{eq:approx} can be now written as
\begin{equation} \label{eq:approx_2}
	\gamma_{\rm{KHI}}=\frac{k_{z}^2U_{\rm{0}}^2}{\bar{\nu}}=\frac1{\bar{\nu}}\frac{4\pi^2}{\lambda^2}U_{\rm{0}}^2 \Rightarrow \bar{\nu}=\frac{4\pi^2}{\lambda^2}\frac{U_{\rm{0}}^2}{\gamma_{\rm{KHI}}}.
\end{equation}
Values of the three parameters that appear in the right hand side of Eq. \eqref{eq:approx_2}, namely the flow velocity, the perturbation wavelength and the KHI growth rate, could be estimated from observations. Consequently, through this formula, we could estimate the coupling degree between the two components of the plasma. 

\section{Conclusions} \label{sec:conclusions}
Although they may not embrace all the physics of the considered system, simple models like the one developed in this paper allow focusing on a particular effect and ease the interpretation of the results. 

Here, we have studied how the existence of a neutral component in a plasma affects the propagation of waves and the possible occurrence of KHI in a cylindrical magnetic flux tube. We used a two-fluid theory to obtain the dispersion relation for linear incompressible waves. Then, we studied the dependence of the solutions on several physical parameters, namely the shear flow velocity, the collision frequency between ions and neutrals, the longitudinal wavenumber and the azimuthal wavenumber. We have found that perturbations at an interface separating two partially ionized plasmas are unstable for any velocity shear, contrary to what occurs in fully ionized plasmas, in which the effect of the magnetic field only allows the onset of the KHI for super-Alfvénic shear flows. Perturbations with large longitudinal wavenumbers have higher growth rates than those with smaller values of that parameter. The two constituent fluids of the plasma, i.e., neutral and ionized, are coupled through collisions. This coupling holds a significant influence: increasing the ion-neutral collision frequency reduces the growth rates of the unstable perturbations, although it is not possible to avoid the onset of the KHI. These results are consistent with previous works like the ones developed in simpler configurations by, e.g., \citet{2004ApJ...608..274W} and \citet{2012ApJ...749..163S}. Eq. \eqref{eq:dis_rel} of the present paper reduces to equation (37) of \citet{2012ApJ...749..163S} or equation (26) of \citet{2004ApJ...608..274W} in the TT limit, i.e., when the product of the longitudinal wavenumber by the radius of the tube is much lower than one. Moreover, in the absence of an equilibrium flow and when the densities of neutrals go to zero, we can recover from our dispersion relation the incompressible limit of equation (8a) of \citet{1983SoPh...88..179E}, corresponding to a fully ionized cylindrical flux tube.

We have also found that the solutions of the dispersion relation slightly depend on the azimuthal wavenumber. Modes with a higher azimuthal wavenumber are more unstable, but this variation is only appreciable beyond the TT limit and it has not a great significance.

Then, we applied our model to a thread of a quiescent prominence, by choosing values for the densities and temperatures typical of those coronal features and solving the dispersion relation for several degrees of ionization and wavelengths. We have concluded that, for a certain combination of parameters, the turbulent flows detected in quiescent prominences may be interpreted as consequences of KHI in partially ionized plasmas. The growth rates of the instability increase when rising the ionization fraction, i.e., when the relative densities of neutrals increase.
 
Furthermore, we have provided an analytical approximation of the KHI growth rates for slow shear flows and strong ion-neutral collisional coupling. This formula, Eq. \eqref{eq:approx}, is easier to handle than the full dispersion relation and thus easier to interpret: growth rates of the KHI show a quadratic dependence on the longitudinal wavenumber and the shear flow velocity, and are inversely proportional to the ion-neutral collision frequency. The analytical approximation may be useful in the field of prominence seismology. From it we can define a mean collision frequency, $\bar{\nu}$ (given by Eq. \eqref{eq:mean_col}), that provides an estimation of the coupling degree of the plasmas. Values of this parameter may be computed from observational data.

And even though we have used a model where the magnetic field is equal in both media, the conclusions extracted from its analysis may also be valid when the magnetic field inside the flux tube differs from that outside. 

For the sake of simplicity, in our model we have ignored effects like gravity, compressibility or surface tension. We have also chosen a particular alignment between the mass flow and the magnetic field: they are parallel, a configuration that gives a higher stability to the system. In future investigations the model used here can be improved including the mentioned effects, that may have an impact on the KHI. Moreover, we have focused only on the linear phase of the instability, while the non-linear regime would be also of interest. And we have assumed that the plasma is composed only of hydrogen, while actually more elements with several states of ionization are involved. A much more complex model that incorporates various of the mentioned refinements would be needed to fully understand the KHI in prominences, but such model could not be studied analytically and numerical simulations would be required. Our paper is an improvement from previous investigations but there is space for a lot of future work in this field.

\begin{acknowledgements} 
	We thank the anonymous referee for helpful remarks and suggestions. We acknowledge the financial support from the Spanish MINECO through project AYA2011-22846, from CAIB through the "Grups Competitius" program, and from FEDER funds. DM acknowledges support from MINECO through a "FPI" grant and through funding for a short stay at KU Leuven. In addition, DM acknowledges the hospitality of the Centre for mathematical-Plasma Astrophysics of KU Leuven, and specially thanks Tom Van Doorsselaere and Marcel Goossens for their supervision and advice during his short stay. RS acknowledges discussion with the ISSI teams on "Partially ionized plasmas in astrophysics” and “Implications for coronal heating and magnetic fields from coronal rain observations and modelling" and thanks ISSI for their support. RS also acknowledges support from MINECO through a "Juan de la Cierva" grant, from MECD through project CEF11-0012, and from the "Vicerectorat d'Investigació i Postgrau" of the UIB. JT acknowledges support from the Spanish Ministerio de Educación y Ciencia through a "Ramón y Cajal" grant.
\end{acknowledgements}

\bibpunct{(}{)}{;}{a}{}{,} %to follow the A&A bib style
\bibliographystyle{aa}
\bibliography{bib_KHI}

%\appendix \label{appendix1}
%\section{}

\end{document}